\documentclass[%
twocolumn,
superscriptaddress,
 amsmath,amssymb,
 aps,
floatfix,
]{revtex4}

\usepackage{graphicx}% Include figure files
\usepackage{dcolumn}% Align table columns on decimal point
\usepackage{bm}% bold math
\usepackage[mathlines]{lineno}% Enable numbering of text and display math
\usepackage[normalem]{ulem}

\usepackage{amssymb}
\usepackage{verbatim} %%remove this later
\usepackage{gensymb} %% for degrees/angle
\usepackage{soul} % for strikethrough
\usepackage{etoolbox} % provides \patchcmd
\usepackage[dvipsnames]{xcolor}

\newcommand{\ket}[1]{\left|  #1 \right\rangle}

\newcommand{\aver}[1]{\ensuremath{\langle {#1} \rangle}}

\newcommand{\var}[1]{\ensuremath{\left( \Delta #1 \right)^2}}

\definecolor{plotgreen}{RGB}{0,150,0}

\begin{document}

\title{Entanglement-Enhanced Optical Atomic Clock}

\author{Edwin Pedrozo-Pe\~{n}afiel}%
\thanks{These authors contributed equally}

\affiliation{Department of Physics, MIT-Harvard Center for Ultracold Atoms and Research Laboratory of Electronics, Massachusetts Institute of Technology, Cambridge, Massachusetts 02139, USA}

\author{Simone Colombo}%
\thanks{These authors contributed equally}
\affiliation{Department of Physics, MIT-Harvard Center for Ultracold Atoms and Research Laboratory of Electronics, Massachusetts Institute of Technology, Cambridge, Massachusetts 02139, USA}

\author{Chi Shu}%
\thanks{These authors contributed equally}
\affiliation{Department of Physics, MIT-Harvard Center for Ultracold Atoms and Research Laboratory of Electronics, Massachusetts Institute of Technology, Cambridge, Massachusetts 02139, USA}
\affiliation{Department of Physics, Harvard University, Cambridge, Massachusetts 02138, USA}

\author{Albert F. Adiyatullin}%
\affiliation{Department of Physics, MIT-Harvard Center for Ultracold Atoms and Research Laboratory of Electronics, Massachusetts Institute of Technology, Cambridge, Massachusetts 02139, USA}

\author{Zeyang Li}%
\affiliation{Department of Physics, MIT-Harvard Center for Ultracold Atoms and Research Laboratory of Electronics, Massachusetts Institute of Technology, Cambridge, Massachusetts 02139, USA}

\author{Enrique Mendez}%
\affiliation{Department of Physics, MIT-Harvard Center for Ultracold Atoms and Research Laboratory of Electronics, Massachusetts Institute of Technology, Cambridge, Massachusetts 02139, USA}

\author{Boris Braverman}
\altaffiliation{Current address: Department of Physics and Max Planck Centre for Extreme and Quantum Photonics, University of Ottawa, 25 Templeton Street, Ottawa, Ontario K1N 6N5, Canada}
\affiliation{Department of Physics, MIT-Harvard Center for Ultracold Atoms and Research Laboratory of Electronics, Massachusetts Institute of Technology, Cambridge, Massachusetts 02139, USA}

\author{Akio Kawasaki}
\altaffiliation{Current address: W. W. Hansen Experimental Physics Laboratory and Department of Physics, Stanford University, Stanford, California 94305, USA}
\affiliation{Department of Physics, MIT-Harvard Center for Ultracold Atoms and Research Laboratory of Electronics, Massachusetts Institute of Technology, Cambridge, Massachusetts 02139, USA}

\author{Daisuke Akamatsu}
\affiliation{Department of Physics, MIT-Harvard Center for Ultracold Atoms and Research Laboratory of Electronics, Massachusetts Institute of Technology, Cambridge, Massachusetts 02139, USA}
\affiliation{National Metrology Institute of Japan (NMIJ), National Institute of Advanced Industrial Science and Technology (AIST), 1-1-1 Umezono, Tsukuba, Ibaraki 305-8563, Japan}

\author{Yanhong Xiao}
\affiliation{Department of Physics, MIT-Harvard Center for Ultracold Atoms and Research Laboratory of Electronics, Massachusetts Institute of Technology, Cambridge, Massachusetts 02139, USA}
\affiliation{State Key Laboratory of Quantum Optics and Quantum Optics Devices, Institute of Laser Spectroscopy, and Collaborative Innovation Center of Extreme Optics, Shanxi University, Taiyuan, Shanxi 030006, China}

\author{Vladan Vuleti\'{c} }
\email{vuletic@mit.edu}
\affiliation{Department of Physics, MIT-Harvard Center for Ultracold Atoms and Research Laboratory of Electronics, Massachusetts Institute of Technology, Cambridge, Massachusetts 02139, USA}

\date{\today}

\begin{abstract}
State-of-the-art atomic clocks are based on the precise detection of the energy difference between two atomic levels, measured as a quantum phase accumulated in a given time interval \cite{Ludlow2015,Ushijima2015,oelker2019demonstration,Schioppo2017Ultrastable}. Optical-lattice clocks (OLCs) now operate at or near the standard quantum limit (SQL) that arises from the quantum noise associated with discrete measurement outcomes.
While performance beyond the SQL has been achieved in microwave clocks and other atomic sensors by engineering quantum correlations (entanglement) between the atoms \cite{Appel2009,Takano2009,gross2010nonlinear,Riedel2010,Leroux2010,kruse2016improvement,Pezze2018,Cox2016a,Hosten2016,Bohnet2016,braverman2019near}, the generation of entanglement on an optical-clock transition and operation of such a clock beyond the SQL represent major goals in quantum metrology that have never been demonstrated \cite{Pezze2018}.
Here we report creation of a many-atom entangled state on an optical transition, and demonstrate an OLC with an Allan deviation below the SQL.
We report a metrological gain of $4.4_{-0.4}^{+0.6}~\mathrm{dB}$ over the SQL using an ensemble consisting of a few hundred $^{171}$Yb atoms, allowing us to reach a given stability $2.8 \pm 0.3$ times faster than the same clock operated at the SQL.
Our results should be readily applicable to other systems, thus enabling further advances in timekeeping precision and accuracy. Entanglement-enhanced OLCs will have many scientific and technological applications, including precision tests of the fundamental laws of physics \cite{Wcislo2018,Safronova2018RevModPhys, Safronova2019}, geodesy \cite{lisdat2016clock, grotti2018geodesy,Katori2020}, or gravitational wave detection \cite{Kolkowitz2016}.

\end{abstract}
% \pacs{03.65.Aa, 03.67.Bg, 32.80.Qk}% PACS, the Physics and Astronomy
\let\clearpage\relax                              
\maketitle

%\section{Introduction}
%
Progress in atomic, optical, and quantum physics over the last decades has boosted the performances of OLCs to an astonishing fractional stability at the level of $10^{-18}$~\cite{Ludlow2015,Ushijima2015,oelker2019demonstration,Schioppo2017Ultrastable}.  Recently, technical noise in OLCs has been reduced to near or below the level of the intrinsic quantum noise \cite{Schioppo2017Ultrastable, oelker2019demonstration}. A clock operated with $N$ uncorrelated atoms for an averaging time $T$ at the SQL can reach a quantum-noise-limited fractional stability given by
\begin{equation}
    \sigma(\tau_R,T)=\frac{1}{\omega_0 \tau_R}\sqrt{\frac{T_C}{T}}\sqrt{\frac{\xi_W^2}{N }},
 \label{eq:ClockPrecision}
\end{equation}
where $\tau_R$ is the interaction time of the atoms with the clock laser (local oscillator, LO),
$T_C$ the clock cycle time, $\omega_0$ the angular frequency of the clock transition, and the Wineland parameter is $\xi^2_W{=}1$~\cite{Wineland1994} for ideal conditions with perfect quantum coherent state preparation and detection. If the clock is operated at duty cycle less than 1 ($T_C{>}\tau_R$), so that the LO is not locked to the atomic evolution during part of the cycle, then the Dick noise $\sigma_\mathrm{Dick}^2$, arising from aliased high-frequency noise of the LO, should be added to Eq. \ref{eq:ClockPrecision}~\cite{dick1987local, Norcia2019}. However, the Dick noise can be suppressed by using two ensembles to eliminate the dead time \cite{Schioppo2017Ultrastable, Norcia2019}, or by performing simultaneous interrogation of two ensembles \cite{takamoto2011frequency, Nicholson2012}.
%

%%% ---------- Fig 1
\begin{figure*}[h!t]
\setlength{\unitlength}{1\textwidth}
\includegraphics[width=143mm,scale=1]{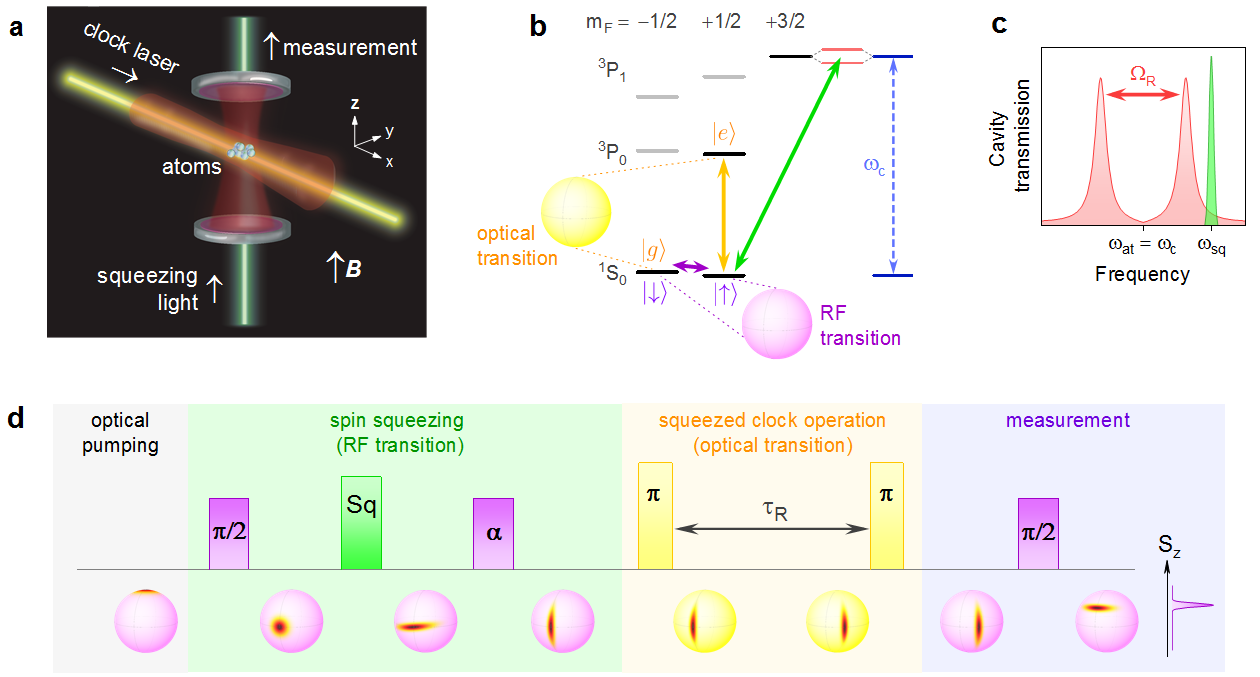}
\caption{
\textbf{Setup and squeezed-clock sequence.} 
\textbf{a}, ${}^{171}$Yb atoms are trapped inside an optical cavity in a two-dimensional magic-wavelength optical lattice along the $x$ and $z$ directions (red beams). Light for optical pumping and spin squeezing (green) is applied along the cavity axis $z$, while the clock laser (yellow) propagates along $x$. 
\textbf{b}, Energy levels and transitions. Purple, green, and yellow pulses indicate the ground-state radiofrequency (RF) transition $\ket{\downarrow} \rightarrow \ket{\uparrow}$, squeezing transition $\ket{{}^1S_0, m_I=\frac{1}{2}} \rightarrow \ket{{}^3P_1, m_F=\frac{3}{2}}$ at $\omega_{at}$, and optical-clock transition $\ket{\uparrow}\rightarrow\ket{e}$, respectively.
The system evolves either in the ground-state manifold $\{ \ket{\downarrow}, \ket{\uparrow} \}$ (purple Bloch sphere), or in the clock-state manifold $\{\ket{g},\ket{e}\}$ (yellow Bloch sphere). The cavity frequency $\omega_c$ is tuned in resonance with the squeezing transition, $\omega_c=\omega_{at}$.
\textbf{c}, Spin squeezing. Strong coupling of the atoms to the cavity results in vacuum Rabi splitting of the cavity resonance (red peaks). A laser is applied detuned from the Rabi peak at frequency $\omega_{sq}$ (green peak) to produce a SSS via cavity feedback \cite{braverman2019near}.
\textbf{d}, Squeezed-clock sequence. A SSS is prepared in the ground-state manifold $\{\ket{\downarrow}, \ket{\uparrow} \}$, transferred to the clock manifold $\{ \ket{g},\ket{e}  \}$, evolved in a Ramsey sequence for time $\tau_R$, and mapped back onto $\{\ket{\downarrow}, \ket{\uparrow} \}$, where a state measurement is performed. 
The evolution of the quantum state is depicted on the Bloch spheres for the RF (purple) and optical (yellow) transitions. 
}
\label{fig:Scheme}
\end{figure*}

The SQL, as described by Eq. \ref{eq:ClockPrecision} with $\xi^2_W{=}1$, is not a fundamental limit, but can be overcome by means of quantum correlations (entanglement) between the participating atoms. The simplest entangled state offering metrological gain is the squeezed spin state (SSS)~\cite{Kitagawa1993}, where the quantum noise is redistributed between two orthogonal spin quadratures, one with reduced quantum noise (squeezed axis), and the other with increased  noise (anti-squeezed axis), see Fig.~\ref{fig:Scheme}d. In this picture, each atom is associated with a spin $\frac{1}{2}$, and the $N$-atom ensemble with a collective spin $S_0=N/2$. By orienting the squeezed quadrature of the collective spin along the phase axis during clock operation one can reduce the quantum noise and increase the clock precision. The metrological gain over the SQL, expressed in variance, is then given by $\xi_W^{-2}$, where the Wineland parameter $\xi_W^{2}=\xi^{2}/C^{2}$~\cite{Wineland1994} incorporates both the the variance reduction $\xi^{2}$ of the spin noise compared to an unentangled coherent spin state (CSS), and the mean spin vector length $\aver{|\vec{S}|}=C S_0$.

Over the last decade, SSSs have been demonstrated in several systems \cite{Pezze2018}, including atomic Bose-Einstein condensates~\cite{Riedel2010, gross2010nonlinear, Hamley2012, kruse2016improvement}, cold atomic ensembles~\cite{Takano2009, Appel2009,Leroux2010, Schleier-Smith2010, Cox2016a, braverman2019near}, and trapped ions ~\cite{Bohnet2016}. In neutral atoms, up to 20~dB of spin squeezing beyond the SQL has been demonstrated using optical techniques~\cite{Hosten2016,Cox2016a}. However, given that it is more difficult to maintain phase coherence at high frequencies, all spin squeezing so far has involved transitions whose frequencies $\omega_0$ are five to ten orders of magnitude smaller than optical frequencies, and that exhibit proportionally reduced time-keeping precision (see Eq. \ref{eq:ClockPrecision}). Building on spin squeezing generation between Zeeman sublevels of the electronic ground state of $^{171}$Yb that we have recently demonstrated~\cite{braverman2019near}, here we report the generation of a SSS on an optical transition, and demonstrate, for the first time, an OLC with performance beyond the SQL.

%% ---- \section{Set up and Experiment description}

Our clock operates with an ensemble of $N=350 \pm 40$ $^{171}\mathrm{Yb}$ atoms that are confined in a two-dimensional magic-wavelength optical-lattice trap inside a high-finesse ($\mathcal{F}{\approx}12000$) optical cavity \cite{braverman2019near} (see Fig. \ref{fig:Scheme}a), and are Raman sideband cooled to mean vibration quantum number $\aver{n_x}<0.2$.
In order to robustly create a SSS on the ultra-narrow optical-clock transition, we first generate spin squeezing between the two nuclear sublevels $\ket{\uparrow} \equiv \ket{^1S_0, m_I=+\frac{1}{2}}$ and $\ket{\downarrow} \equiv \ket{^1S_0, m_I=-\frac{1}{2}}$ of the electronic ground state $^1S_0$ using the interaction between the atoms and the optical cavity \cite{Leroux2010,braverman2019near}. Subsequently we transfer the population of $\ket{\uparrow}$ into the $\ket{e} \equiv \ket{^3P_0,m_I=\frac{1}{2}}$ excited clock state with a $\pi$ pulse of the clock laser, thereby mapping the SSS onto the optical-clock manifold $\{\ket{g}{\equiv} \ket{\downarrow},\ket{e}\}$ (see~Fig. \ref{fig:Scheme}b).

The spin squeezing between the ground-state sublevels is achieved by optically pumping the atoms into state $\ket{\uparrow}$, creating a CSS between $\ket{\uparrow}$ and $\ket{\downarrow}$ with a radiofrequency (RF) $\pi/2$ pulse, and then applying a laser pulse near the $\ket{\uparrow} \rightarrow \ket{^3P_1,m_F=\frac{3}{2}}$ transition through the cavity~ \cite{braverman2019near} (Fig. \ref{fig:Scheme}c). The atom-light interaction, amplified by the cavity, approximates the one-axis twisting Hamiltonian \cite{Kitagawa1993} with longitudinal magnetic field $H_1 = \beta\,S_z+\chi\,S_z^2$.
A spin-echo protocol is then used to cancel the linear term ($S_z$), such that the system evolves under an effective one-axis twisting Hamiltonian $H{=}\chi\,S_z^2$~\cite{Kitagawa1993, braverman2019near} for a time $\tau_s$ (see Methods and Fig. \ref{fig:Scheme}d). Finally, the reduced spin projection noise can be oriented along any desired axis in the $\{\ket{\uparrow},\ket{\downarrow} \}$ ground-state manifold by rotating the SSS around its average spin direction $\aver{\vec{S}}$ with another RF pulse. At this stage, we observe a spin noise suppression $\xi^{-2} =6$~dB (limited by the state detection~(see Methods), with an intrinsic spin noise reduction of 9~dB), and a contrast of $C=0.97$. The SSS is nearly uncertainty limited, with a detected (inferred) area 1.9 (1.6) times larger than the limit set by the Heisenberg's uncertainty principle \cite{braverman2019near}.

\begin{figure}[t!p]
\setlength{\unitlength}{1\textwidth}
\includegraphics[width=79mm,scale=.9]{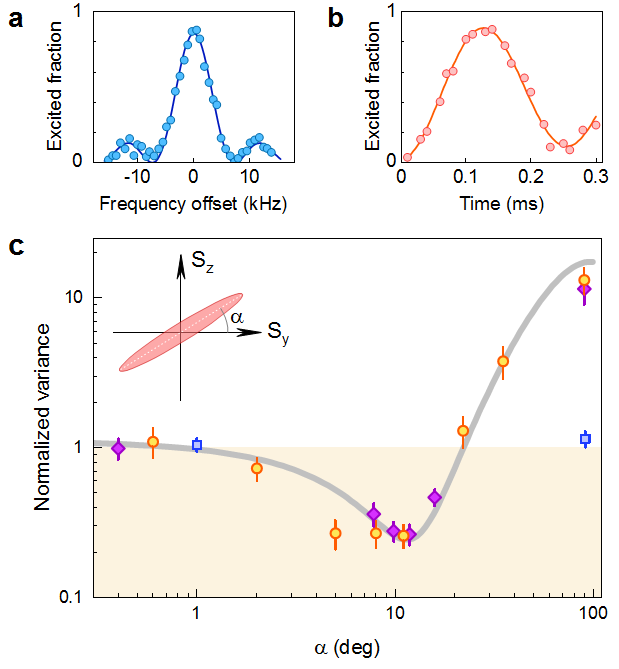}

\caption{
\textbf{Squeezed state tomography.} 
\textbf{a}, Rabi spectroscopy of the $n_x \rightarrow n_x$ vibrational component of the $\ket{\uparrow} \rightarrow \ket{e}$ transition to the clock state. The pulse duration is 0.22~ms. \textbf{b}, Rabi oscillations on the clock transition.
\textbf{c}, Tomography measurements of the SSS before (purple diamond) and after (orange circle) mapping onto the clock transition and back show very similar degrees of squeezing. The theoretical expectation for a state with $\xi^2_{-}=-5.9$ dB is shown as a gray solid line.
Two measured data points for a CSS are also shown (blue squares).
The shaded area indicates the region below the SQL.
In this and in following figures, error bars represent the $68\%$ confidence interval.
}
\label{fig:tomography}
\end{figure}

Having prepared a SSS in the $\{\ket{\uparrow},\ket{\downarrow} \}$ ground-state manifold, we then map it onto the optical-clock transition ${}^1S_0 \rightarrow {}^3P_0$ by phase-coherently transferring the population of $\ket{\uparrow}$ to the state $\ket{e}$ with  an optical $\pi$ pulse of the LO.
We observe a clean Rabi spectrum on the $\ket{g} \rightarrow \ket{e}$ transition (Fig. \ref{fig:tomography}a), and coherent Rabi oscillations in time (Fig. \ref{fig:tomography}b). The $\pi$-pulse has a transfer efficiency of 0.95, setting a limit on the observable noise reduction of 13~dB below the projection noise limit, being well below the spin noise of our SSS.
Thus we do not expect any substantial degradation of the spin squeezing to arise from mapping the SSS onto the $\ket{g} \rightarrow \ket{e}$ clock transition.

\begin{figure}[hbtp]
\setlength{\unitlength}{1\textwidth}
\includegraphics[width=79mm,scale=.9]{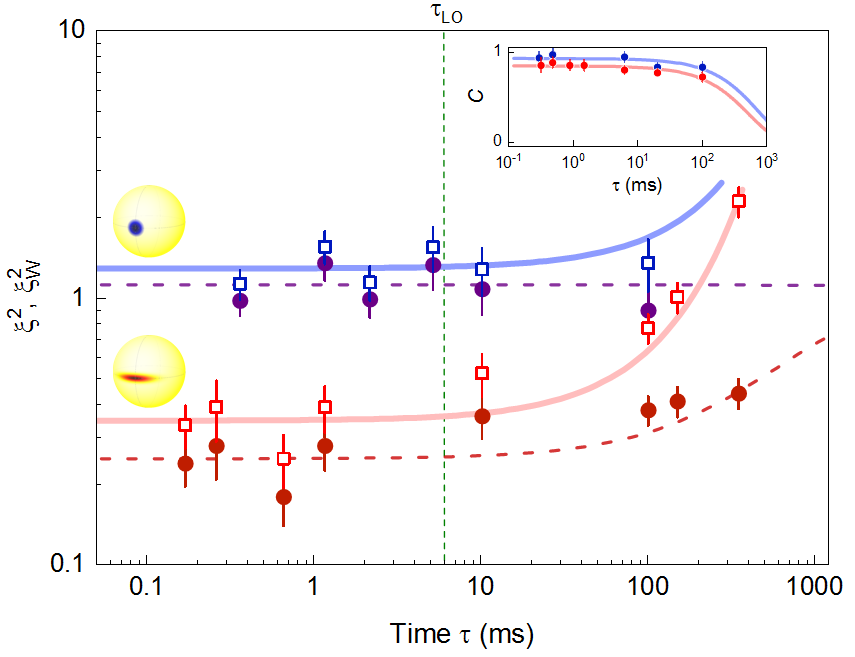}
\caption{\textbf{Spin noise and Wineland parameter on the clock transition as a function of time.}
The measured normalized $S_z$ spin noise $\xi^2$ (solid circles) and model (dashed lines) for the CSS (purple) and the SSS with squeezed axis aligned along $S_z$ (dark-red) increases slowly with time. The increase is caused by the finite 800-ms lifetime of the state $\ket{e}$ due to scattering of the trapping light. The Wineland parameter $\xi_W^2$ (open squares) depends in addition on the contrast loss due to shortening of the atomic collective spin $\aver{|\vec{S}|}$ (inset, blue data points for CSS, red for SSS, with exponential fits). The solid lines represent the predicted Wineland parameter using the fit to the measured contrast. $\tau_{LO}$ indicates the coherence time of the LO.}
\label{fig:squeezingSurvival}
\end{figure}

We first demonstrate that the entanglement survives the transfer $\ket{g} \rightarrow \ket{e} \rightarrow \ket{g}$ by characterizing the spin squeezing remaining after this process in the $\{ \ket{\uparrow}, \ket{\downarrow} \}$ manifold via state tomography (Fig.~\ref{fig:tomography}c). We observe that the squeezed noise dips significantly below the standard quantum limit to a level $\xi^2=-5.9_{-0.8}^{+0.6}\,\mathrm{dB}$, and is essentially the same before and after the mapping onto the clock levels.
In order to determine the entanglement-induced metrological gain $\xi_W^{-2}$ available on the clock transition, we measure the contrast $C$ for a Ramsey sequence composed of the preparation of a CSS or SSS within the $\{ \ket{\uparrow}, \ket{\downarrow} \}$ ground-state manifold, subsequent optical $\pi$-pulses $\ket{\uparrow} \rightarrow \ket{e}$ separated by a Ramsey dark time $\tau_R =0.23$~ms, and a final ground state RF $\pi{/}2$-pulse followed by a measurement of $S_z$ in the $\{ \ket{\uparrow}, \ket{\downarrow} \}$ spin space (see Fig. \ref{fig:Scheme}d). We measure $C=0.85 \pm 0.01$, yielding a Wineland parameter $\xi_W^{2}{=}-4.4_{-0.6}^{+0.4}\,\mathrm{dB}$ for the SSS in the optical manifold $\{\ket{g},\ket{e}\}$, and corresponding to a potential reduction in averaging time to reach a given precision by a factor $2.8 \pm 0.3$.

%%%-------------FIG: 4
\begin{figure*}[h!t]
\setlength{\unitlength}{1\textwidth}
\includegraphics[width=153mm,scale=.9]{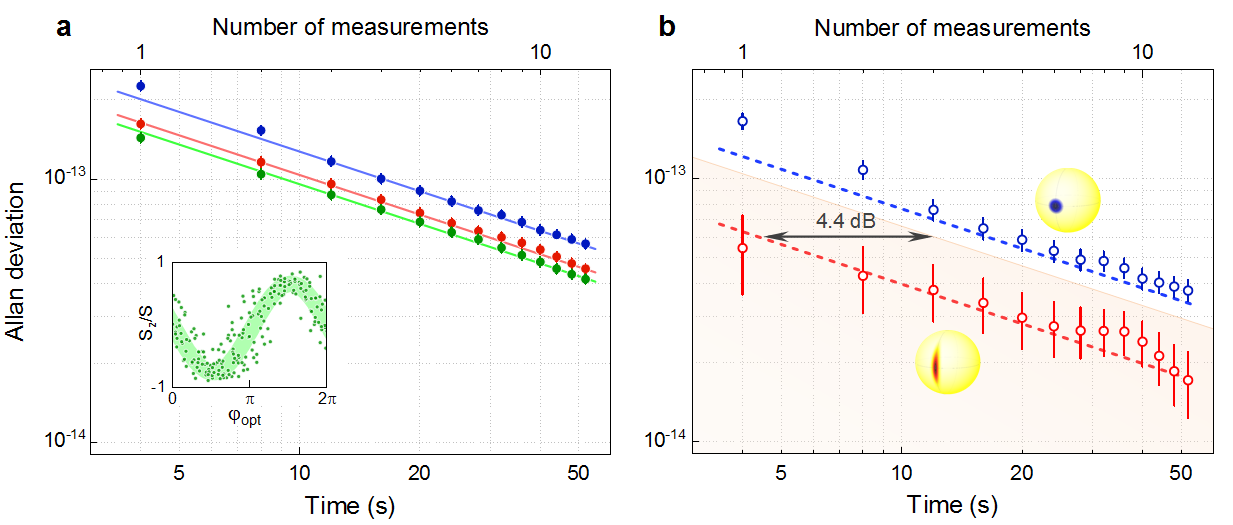}
\caption{\textbf{Stability improvement with the squeezed clock.} 
\textbf{a}, Self-comparison Allan deviation~\cite{Nicholson2012,Bloom2014} for a CSS with Ramsey time $\tau_{\mathrm{R1}}=0.17$~ms (sequence \emph{C1}, blue), a SSS state with $\tau_{\mathrm{R1}}=0.17$~ms (\emph{R1}, red), and a SSS with $\tau_{\mathrm{R2}}=1.16$~ms (\emph{R2}, green).
Inset: Ramsey fringes vs. LO phase $\phi_{opt}$ of the second Ramsey $\pi-$pulse for \emph{R2}.
\textbf{b}, Data for  \emph{C1} and \emph{R1} after subtraction of the phase noise of the LO, calculated from the data for \emph{R2}.
The dashed lines indicate the expected performance estimated from the separately measured Wineland parameter $\xi^2_W$, see Fig.~\ref{fig:squeezingSurvival}. The shaded area represents the region below the SQL. The other parameters for these measurement are: atom number $N{=}300{\pm}30$, cycle time $T_C{=}4~$s, Ramsey contrast $C_{C1}=0.91(1)$ for CSS, $C_{R1}=0.85(1)$ and $C_{R2}=0.70(1)$ for SSS.}
\label{fig:Allan}
\end{figure*}

Figure~\ref{fig:squeezingSurvival} shows the measured normalized spin noise $\xi^{2}$ and Wineland parameter $\xi_W^{2}$ as a function of Ramsey time on the optical-clock transition for both a CSS, and for a SSS with its squeezed direction oriented along $S_z$ in the $\{\ket{g},\ket{e}\}$ spin space.
In this configuration, that does not improve clock performance, but can be used to characterize the entanglement \cite{leroux2010orientation}, the LO phase noise does not affect the squeezed quadrature $S_z$, and we observe that the $S_z$ spin noise remains reduced for times as long as 1~s. While the observed Ramsey contrast decays due to LO phase noise with a time constant of $\tau_\mathrm{LO}=6$~ms, the intrinsic coherence of the atomic state (mean atomic spin vector length $\aver{|\vec{S}|}$) can be determined even when the LO phase noise is dominant (see Methods), and the corresponding result is displayed for both the CSS and the SSS in the inset to Fig.~ \ref{fig:squeezingSurvival}. The ensemble spin coherence, in the optical-clock manifold, decays exponentially with a time constant $\tau_{\text{ens}}=0.8 \pm 0.2$~s both for the CSS and the SSS. As Fig. \ref{fig:squeezingSurvival} shows, after an interaction time of $\approx 0.2$~s the SSS is no longer sufficiently entangled to overcome the SQL, but it can still offer metrological gain over the CSS for up to 0.5~s.

In a fully operating atomic clock, the atomic phase is used to stabilize the LO phase through feedback.
Hence clock performance can be improved by the use of SSSs with the squeezed axis oriented along the phase direction, which allow measuring the phase difference between atoms and LO with higher precision than a CSS, and apply corresponding feedback \cite{wineland1998experimental}. The optimum Ramsey time that achieves maximum gain over the CSS is determined by the combination of LO coherence time $\tau_\mathrm{LO}$ and Wineland parameter \cite{Braverman2018}. 

In Fig. \ref{fig:Allan}a we demonstrate directly that an OLC using an entangled state can perform below the SQL. We implement the full clock sequence depicted in Fig. \ref{fig:Scheme}d with Ramsey times of $\tau_{\mathrm{R}1}=0.17$~ms (Ramsey sequence \emph{R1}) and $\tau_{\mathrm{R}2}=1.16$~ms (Ramsey sequence \emph{R2}). $\tau_{\mathrm{R}2}$ is chosen so that the measured phase noise between atoms and light is dominated by the LO noise (see Methods). 
Since the cycle time $T_C=4~s \gg \tau_{\mathrm{R1}},\tau_{\mathrm{R}2}$, it is reasonable to assume the LO noise is the same in both sequences \emph{R1, R2} (see Methods). We can then remove the LO noise, as measured in \emph{R2}, from the Allan deviation of the \emph{R1} data, and thus obtain the intrinsic stability of the clock operated with the SSS in \emph{R1} (red circles in Fig.~\ref{fig:Allan}b). We further show for comparison a clock operated with a CSS under the same conditions, and with LO noise also removed (blue circles). The Allan deviation of the squeezed clock is visibly lower. 

From the Wineland parameter obtained in Fig.~\ref{fig:squeezingSurvival} $\xi_W^2{=}-4.4_{-0.6}^{+0.4}$~dB, we expect a squeezed-clock Allan deviation of ${\sigma_{sq}=(1.25\pm0.10){\times}10^{-13}~\mathrm{s^{1/2}{/}}\sqrt{T}}$ (red dashed line in Fig:~\ref{fig:Allan}b), in excellent agreement with the data at ${\sigma{=}(1.30\pm0.04) \times10^{-13}~\mathrm{s^{1/2}{/}}\sqrt{T}}$. The corresponding SQL for a perfect clock operating with a CSS (with the same cycle time, Ramsey time, atom number, and perfect state detection), as given by Eq. \ref{eq:ClockPrecision}, is $\sigma_\mathrm{SQL}{=}2.1\times10^{-13}~\mathrm{s^{1/2}{/}}\sqrt{T}$. Compared to the CSS, the improvement is larger, $5.7~$dB, corresponding to a factor 3.7 reduction in averaging time. This is the case because the CSS is also subject to imperfect state detection and contrast loss.

In conclusion, we have generated for the first time spin squeezing for many atoms between atomic levels whose energy differs on the scale of optical photons.
We have used this entanglement to directly demonstrate a squeezed OLC performing $4.4_{-0.4}^{+0.6}$~dB beyond the SQL, corresponding to a $2.8{\pm}0.3$ shorter averaging time.
In the future, this system can be improved in several ways: Using a LO with much longer coherence time, we would be limited by the atomic decoherence due to scattering of the trap light~(Fig.~\ref{fig:squeezingSurvival}). This effect can be mitigated by reducing the trap-light intensity.
Furthermore, the amount of metrological gain demonstrated in our system is primarily limited by the state detection noise that accounts for almost half of the total detected spin variance~\cite{braverman2019near}.
With improved state detection and for an increased atom number of $N{=}1000$ we can reach a Wineland parameter $\xi_W^2{=}-13~$dB~\cite{braverman2019near}, which already exceeds the optimum $\xi_W^2{\approx}-12$~dB beyond which the effective metrological gain degrades with increased squeezing~\cite{Braverman2018}.
Then, assuming an interaction time $\tau_R{=}300~$ms and a $50\%$ duty cycle, we would reach a quantum-noise-limited fractional stability of $\sigma_{sq}{=}1\times 10^{-17}~\mathrm{s^{1/2}{/}}\sqrt{T}$.
Such stability would allow one, e.g., to measure a gravitational redshift for a height difference of 1~cm in less than 2~minutes.

%% ----- Acknowledgements ----- %%
\begin{acknowledgments}

We would like to thank Hidetoshi Katori, Wolfgang Ketterle, Andrew Ludlow, Mikhail Lukin, Joshua Ramette, Giacomo Roati, Alban Urvoy, Zachary Vendeiro, and Jun Ye for discussions. This work was supported by NSF, DARPA, ONR, and the NSF Center for Ultracold Atoms (CUA). S.C. and A.A. acknowledge support from the Swiss National Science Foundation (SNSF). B.B. acknowledges the support of the Banting Postdoctoral Fellowship. A.K. acknowledges the partial support of a William M. and Jane D. Fairbank Postdoctoral Fellowship of Stanford University.

\end{acknowledgments}

\section*{Author contributions}
A.K., B.B., C.S., E.P.-P., S.C, A.A., Z.L., E.M., and V.V. contributed to the building of the experiment. E.P.-P., S.C., and C.S. led the experimental efforts and simulations. S.C., A.A., C.S., and E.P.-P. contributed to the data analysis. V.V. conceived and supervised the experiment. S.C. and V.V. wrote the manuscript. All authors discussed the experiment implementation, the results, and contributed to the manuscript.

%%----- Bibliography -----%%

\bibliographystyle{ieeetr}
\bibliography{SqOptTransition}{}

\begin{thebibliography}{10}

\bibitem{Ludlow2015}
A.~D. Ludlow, M.~M. Boyd, J.~Ye, E.~Peik, and P.~O. Schmidt, ``Optical atomic
  clocks,'' {\em Rev. Mod. Phys.}, vol.~87, pp.~637--701, Jun 2015.

\bibitem{Ushijima2015}
I.~Ushijima, M.~Takamoto, M.~Das, T.~Ohkubo, and H.~Katori, ``Cryogenic optical
  lattice clocks,'' {\em Nat. Photonics}, vol.~9, pp.~185--189, Mar. 2015.

\bibitem{oelker2019demonstration}
E.~Oelker, R.~Hutson, C.~Kennedy, L.~Sonderhouse, T.~Bothwell, A.~Goban,
  D.~Kedar, C.~Sanner, J.~Robinson, G.~Marti, {\em et~al.}, ``Demonstration of
  $4.8\times10^{-17}$ stability at 1 s for two independent optical clocks,''
  {\em Nature Photonics}, vol.~13, no.~10, pp.~714--719, 2019.

\bibitem{Schioppo2017Ultrastable}
M.~Schioppo, R.~C. Brown, W.~F. McGrew, N.~Hinkley, R.~J. Fasano, K.~Beloy,
  T.~H. Yoon, G.~Milani, D.~Nicolodi, J.~A. Sherman, N.~B. Phillips, C.~W.
  Oates, and A.~D. Ludlow, ``Ultrastable optical clock with two cold-atom
  ensembles,'' {\em Nat. Photonics}, vol.~11, pp.~48--52, Jan. 2017.

\bibitem{Appel2009}
J.~Appel, P.~J. Windpassinger, D.~Oblak, U.~B. Hoff, N.~Kj{\ae}rgaard, and
  E.~S. Polzik, ``Mesoscopic atomic entanglement for precision measurements
  beyond the standard quantum limit,'' {\em Proc. Natl. Acad. Sci. U.S.A.},
  vol.~106, pp.~10960--10965, July 2009.

\bibitem{Takano2009}
T.~Takano, M.~Fuyama, R.~Namiki, and Y.~Takahashi, ``Spin squeezing of a cold
  atomic ensemble with the nuclear spin of one-half,'' {\em Phys. Rev. Lett.},
  vol.~102, p.~033601, Jan 2009.

\bibitem{gross2010nonlinear}
C.~Gross, T.~Zibold, E.~Nicklas, J.~Esteve, and M.~K. Oberthaler, ``Nonlinear
  atom interferometer surpasses classical precision limit,'' {\em Nature},
  vol.~464, no.~7292, pp.~1165--1169, 2010.

\bibitem{Riedel2010}
M.~F. Riedel, P.~B{\"o}hi, Y.~Li, T.~W. H{\"a}nsch, A.~Sinatra, and
  P.~Treutlein, ``Atom-chip-based generation of entanglement for quantum
  metrology,'' {\em Nature (London)}, vol.~464, no.~7292, p.~1170, 2010.

\bibitem{Leroux2010}
I.~D. Leroux, M.~H. Schleier-Smith, and V.~Vuleti\'{c}, ``Implementation of
  cavity squeezing of a collective atomic spin,'' {\em Phys. Rev. Lett.},
  vol.~104, no.~7, p.~073602, 2010.

\bibitem{kruse2016improvement}
I.~Kruse, K.~Lange, J.~Peise, B.~L{\"u}cke, L.~Pezze, J.~Arlt, W.~Ertmer,
  C.~Lisdat, L.~Santos, A.~Smerzi, {\em et~al.}, ``Improvement of an atomic
  clock using squeezed vacuum,'' {\em Physical review letters}, vol.~117,
  no.~14, p.~143004, 2016.

\bibitem{Pezze2018}
L.~Pezz{\`e}, A.~Smerzi, M.~K. Oberthaler, R.~Schmied, and P.~Treutlein,
  ``Quantum metrology with nonclassical states of atomic ensembles,'' {\em Rev.
  Mod. Phys.}, vol.~90, no.~3, p.~035005, 2018.

\bibitem{Cox2016a}
K.~C. Cox, G.~P. Greve, J.~M. Weiner, and J.~K. Thompson, ``Deterministic
  squeezed states with collective measurements and feedback,'' {\em Phys. Rev.
  Lett.}, vol.~116, p.~093602, 2016.

\bibitem{Hosten2016}
O.~Hosten, N.~J. Engelsen, R.~Krishnakumar, and M.~A. Kasevich, ``Measurement
  noise 100 times lower than the quantum-projection limit using entangled
  atoms,'' {\em Nature (London)}, vol.~529, pp.~505--508, Jan. 2016.

\bibitem{Bohnet2016}
J.~G. Bohnet, B.~C. Sawyer, J.~W. Britton, M.~L. Wall, A.~M. Rey, M.~Foss-Feig,
  and J.~J. Bollinger, ``Quantum spin dynamics and entanglement generation with
  hundreds of trapped ions,'' {\em Science}, vol.~352, p.~1297, June 2016.

\bibitem{braverman2019near}
B.~Braverman, A.~Kawasaki, E.~Pedrozo-Pe{\~n}afiel, S.~Colombo, C.~Shu, Z.~Li,
  E.~Mendez, M.~Yamoah, L.~Salvi, D.~Akamatsu, {\em et~al.}, ``Near-unitary
  spin squeezing in {Yb}~171,'' {\em Phys. Rev. Lett.}, vol.~122, no.~22,
  p.~223203, 2019.

\bibitem{Wcislo2018}
P.~Wcis{\l}o, P.~Ablewski, K.~Beloy, S.~Bilicki, M.~Bober, R.~Brown, R.~Fasano,
  R.~Ciury{\l}o, H.~Hachisu, T.~Ido, J.~Lodewyck, A.~Ludlow, W.~McGrew,
  P.~Morzy{\'n}ski, D.~Nicolodi, M.~Schioppo, M.~Sekido, R.~Le~Targat, P.~Wolf,
  X.~Zhang, B.~Zjawin, and M.~Zawada, ``New bounds on dark matter coupling from
  a global network of optical atomic clocks,'' {\em Science Advances}, vol.~4,
  no.~12, 2018.

\bibitem{Safronova2018RevModPhys}
M.~S. Safronova, D.~Budker, D.~DeMille, D.~F.~J. Kimball, A.~Derevianko, and
  C.~W. Clark, ``Search for new physics with atoms and molecules,'' {\em Rev.
  Mod. Phys.}, vol.~90, p.~025008, Jun 2018.

\bibitem{Safronova2019}
M.~S. Safronova, ``The search for variation of fundamental constants with
  clocks,'' {\em Annalen der Physik}, vol.~531, no.~5, p.~1800364, 2019.

\bibitem{lisdat2016clock}
C.~Lisdat, G.~Grosche, N.~Quintin, C.~Shi, S.~Raupach, C.~Grebing, D.~Nicolodi,
  F.~Stefani, A.~Al-Masoudi, S.~D{\"o}rscher, {\em et~al.}, ``A clock network
  for geodesy and fundamental science,'' {\em Nature communications}, vol.~7,
  no.~1, pp.~1--7, 2016.

\bibitem{grotti2018geodesy}
J.~Grotti, S.~Koller, S.~Vogt, S.~H{\"a}fner, U.~Sterr, C.~Lisdat, H.~Denker,
  C.~Voigt, L.~Timmen, A.~Rolland, {\em et~al.}, ``Geodesy and metrology with a
  transportable optical clock,'' {\em Nature Physics}, vol.~14, no.~5,
  pp.~437--441, 2018.

\bibitem{Katori2020}
M.~Takamoto, I.~Ushijima, N.~Ohmae, T.~Yahagi, K.~Kokado, H.~Shinkai, and
  H.~Katori, ``Test of general relativity by a pair of transportable optical
  lattice clocks,'' {\em Nature Photonics}, pp.~1749--4893, 2020.

\bibitem{Kolkowitz2016}
S.~Kolkowitz, I.~Pikovski, N.~Langellier, M.~D. Lukin, R.~L. Walsworth, and
  J.~Ye, ``Gravitational wave detection with optical lattice atomic clocks,''
  {\em Phys. Rev. D}, vol.~94, p.~124043, Dec 2016.

\bibitem{Wineland1994}
D.~J. Wineland, J.~J. Bollinger, W.~M. Itano, and D.~J. Heinzen, ``Squeezed
  atomic states and projection noise in spectroscopy,'' {\em Phys. Rev. A},
  vol.~50, pp.~67--88, Jul 1994.

\bibitem{dick1987local}
G.~J. Dick, ``Local oscillator induced instabilities in trapped ion frequency
  standards,'' tech. rep., CALIFORNIA INST OF TECH PASADENA JET PROPULSION LAB,
  1987.

\bibitem{Norcia2019}
M.~A. Norcia, A.~W. Young, W.~J. Eckner, E.~Oelker, J.~Ye, and A.~M. Kaufman,
  ``Seconds-scale coherence on an optical clock transition in a tweezer
  array,'' {\em Science}, vol.~366, no.~6461, pp.~93--97, 2019.

\bibitem{takamoto2011frequency}
M.~Takamoto, T.~Takano, and H.~Katori, ``Frequency comparison of optical
  lattice clocks beyond the dick limit,'' {\em Nature Photonics}, vol.~5,
  no.~5, p.~288, 2011.

\bibitem{Nicholson2012}
T.~L. Nicholson, M.~J. Martin, J.~R. Williams, B.~J. Bloom, M.~Bishof, M.~D.
  Swallows, S.~L. Campbell, and J.~Ye, ``Comparison of two independent sr
  optical clocks with $1 \times {10}^{-17}$ stability at ${10}^{3}$ s,'' {\em
  Phys. Rev. Lett.}, vol.~109, p.~230801, Dec 2012.

\bibitem{Kitagawa1993}
M.~Kitagawa and M.~Ueda, ``Squeezed spin states,'' {\em Phys. Rev. A}, vol.~47,
  pp.~5138--5143, Jun 1993.

\bibitem{Hamley2012}
C.~D. Hamley, C.~Gerving, T.~Hoang, E.~Bookjans, and M.~S. Chapman,
  ``Spin-nematic squeezed vacuum in a quantum gas,'' {\em Nat. Phys.}, vol.~8,
  no.~4, p.~305, 2012.

\bibitem{Schleier-Smith2010}
M.~H. Schleier-Smith, I.~D. Leroux, and V.~Vuleti\'{c}, ``Squeezing the
  collective spin of a dilute atomic ensemble by cavity feedback,'' {\em Phys.
  Rev. A}, vol.~81, p.~021804(R), Feb 2010.

\bibitem{Bloom2014}
B.~J. Bloom, T.~L. Nicholson, J.~R. Williams, S.~L. Campbell, M.~Bishof,
  X.~Zhang, W.~Zhang, S.~L. Bromley, and J.~Ye, ``An optical lattice clock with
  accuracy and stability at the $10^{-18}$ level,'' {\em Nature (London)},
  vol.~506, no.~7486, pp.~71--75, 2014.

\bibitem{leroux2010orientation}
I.~D. Leroux, M.~H. Schleier-Smith, and V.~Vuleti{\'c}, ``Orientation-dependent
  entanglement lifetime in a squeezed atomic clock,'' {\em Phys. Rev. Lett.},
  vol.~104, no.~25, p.~250801, 2010.

\bibitem{wineland1998experimental}
D.~J. Wineland, C.~Monroe, W.~M. Itano, D.~Leibfried, B.~E. King, and D.~M.
  Meekhof, ``Experimental issues in coherent quantum-state manipulation of
  trapped atomic ions,'' {\em Journal of Research of the National Institute of
  Standards and Technology}, vol.~103, no.~3, p.~259, 1998.

\bibitem{Braverman2018}
B.~Braverman, A.~Kawasaki, and V.~Vuleti\'{c}, ``Impact of non-unitary spin
  squeezing on atomic clock performance,'' {\em New J. Phys.}, vol.~20, no.~10,
  p.~103019, 2018.

\bibitem{Kawasaki2019}
A.~Kawasaki, B.~Braverman, E.~Pedrozo-Pe\~nafiel, C.~Shu, S.~Colombo, Z.~Li,
  O.~\"Ozel, W.~Chen, L.~Salvi, A.~Heinz, D.~Levonian, D.~Akamatsu, Y.~Xiao,
  and V.~Vuleti\ifmmode~\acute{c}\else \'{c}\fi{}, ``Geometrically asymmetric
  optical cavity for strong atom-photon coupling,'' {\em Phys. Rev. A},
  vol.~99, p.~013437, Jan 2019.

\bibitem{Blatt2009PRA}
S.~Blatt, J.~W. Thomsen, G.~K. Campbell, A.~D. Ludlow, M.~D. Swallows, M.~J.
  Martin, M.~M. Boyd, and J.~Ye, ``Rabi spectroscopy and excitation
  inhomogeneity in a one-dimensional optical lattice clock,'' {\em Phys. Rev.
  A}, vol.~80, p.~052703, Nov 2009.

\bibitem{Vallet2017}
G.~Vallet, E.~Bookjans, U.~Eismann, S.~Bilicki, R.~Le~Targat, and J.~Lodewyck,
  ``A noise-immune cavity-assisted non-destructive detection for an optical
  lattice clock in the quantum regime,'' {\em New Journal of Physics}, vol.~19,
  p.~083002, aug 2017.

\bibitem{yamoah2019robust}
M.~Yamoah, B.~Braverman, E.~Pedrozo-Pe{\~n}afiel, A.~Kawasaki,
  B.~Zlatkovi{\'c}, and V.~Vuleti{\'c}, ``Robust khz-linewidth distributed
  bragg reflector laser with optoelectronic feedback,'' {\em Optics Express},
  vol.~27, no.~26, pp.~37714--37720, 2019.

\end{thebibliography}
%% ------ Methods -------%%
\section*{Methods}

\subsection*{Loading of atoms into two-dimensional magical-wavelength optical lattice}

The experimental sequence starts with loading $^{171}$Yb atoms into a two-color mirror magneto-optical trap (MOT) on the singlet ${^1S_0}{\rightarrow} {^1P_1}$ and triplet ${^1S_0}{\rightarrow} {^3P_1}$ transitions, followed by a second-stage green MOT on the triplet transition. By changing the magnetic field, the atomic cloud is then transported into the intersection region of the cavity TEM$_{00}$ mode and a one-dimensional optical lattice along the $x$-direction (see Fig.~\ref{fig:Scheme}). The trap is formed by 'magic-wavelength' light with $\lambda_t\approx759$ nm, and the trap depth is $U_x{=}k_B{\times}10$~$\mu$K. The green MOT light is then turned off, and the magic-wavelength trap inside the cavity, detuned from the $x$ lattice by $160$~MHz to avoid interference, is ramped up in 40 ms to a trap depth $U_{c}{=}k_B{\times}120$~$\mu$K. At the end of loading process, the transverse lattice power is ramped down to zero and back to full power in 50 ms to remove all the atoms that are outside the overlap region of the two lattices. In this way, an ensemble of typically $N=350$ atoms is prepared at a distance of $370$~$\mu$m from the end mirror of the cavity. At this location, the single-atom peak cooperativity is $\eta_0=3.1$ \cite{Kawasaki2019}.

\subsection*{Raman sideband cooling}

After loading the atoms into the two-dimensional optical lattice, the atomic temperature is typically $10$~$\mu$K, as measured by sideband spectroscopy on the clock transition ${^1S_0}{\rightarrow} {^3P_0}$ \cite{Blatt2009PRA}. Subsequently Raman sideband cooling is performed on the transition ${^1S_0}{\rightarrow} {^3P_1}$ in an applied magnetic field $B_z=13.6$~G along the $z$-direction. 
In $100$~ms, the atomic temperature is lowered to $1.8$~$\mu$K, corresponding to an average motional occupation number $\aver{n_x}{=}0.2$ at a trap vibration frequency of $\omega_x/(2\pi)=67$~kHz along the $x$-direction. The cavity trap is then adiabatically ramped down to $U_c{=}k_B{\times}40$~$\mu$K to further reduce temperature and lower the trap light scattering rate on ${^3P_0}$ state.
We observe that during the Raman sideband cooling, where the optical pumping is provided by intracavity light, the atoms reorganize along the lattice such that all atoms have nearly the maximum coupling $\eta_0$ to the cavity mode and the squeezing light.%

\subsection*{State Measurement}
The final measurement of $S_z$ is performed in the ground-state manifold.
It is obtained from the difference $S_z{=}(N_\uparrow{-}N_\downarrow){/}2$ of the populations $N_\uparrow$ and $N_\downarrow$ of the states $\left|\uparrow\right\rangle{=}\left|{^1S_0}; m_F{=}{+}\frac{1}{2}\right\rangle$ and $\left|\downarrow\right\rangle{=}\left|{^1S_0}; m_F{=}{-}\frac{1}{2}\right\rangle$, respectively.
We first measure $N_\uparrow$ through the vacuum Rabi splitting of the cavity mode $2g{\approx}\sqrt{N_\uparrow\eta\kappa\Gamma}$ when the empty cavity is resonant with the transition $\left|\uparrow\right\rangle{\rightarrow}\left|{^3P_1};F{=}3/2,m_F{=}{+}3{/}2\right\rangle$ \cite{Vallet2017,braverman2019near}. Here $\eta{=}3.12(5)$ is the effective single-atom cooperativity, $\kappa{=}2\pi{\times}520(10)~$kHz is the cavity linewidth and $\Gamma{=}2\pi{\times}184(1)~$kHz the linewidth of the atomic transition. The Rabi splitting is measured by scanning the laser frequency and detecting the cavity transmission as a function of the frequency. After that measurement we apply an RF $\pi$-pulse to switch the populations of $\left|\uparrow\right\rangle$ and $\left|\downarrow\right\rangle$. The resulting Rabi splitting of the cavity mode is now proportional to $N_\downarrow$. We then perform a second measurement starting with $N_\downarrow$ to account for atom loss during detection \cite{braverman2019near}, and average the two results according to the formula
$S_z= \left(S_z^{(1)} +  S_z^{(2)} \right)/2= \left(N_\uparrow^{(1)} - N_\downarrow^{(1)} - N_\downarrow^{(2)} + N_\uparrow^{(2)} \right)/4$.
The measurement resolution, expressed in variance normalized to the SQL, is given by  $\sigma_d^2{\equiv}2\mathrm{var}\left(S_z^{(2)}{-}S_z^{(1)}\right)/N$, where $S_z^{(i)}{\equiv}\left(N_\uparrow^{(i)}-N_\downarrow^{(i)}\right)/2$.
Our measurement resolution is $\sigma_d^2{=}0.125$ \cite{braverman2019near}. 
Since all atoms have the same coupling to the cavity, the atom number $N$ inferred from the Rabi splitting equals the real number of atoms in the cavity.

\subsection*{Contrast measurement}

Our goal here is to discriminate loss of coherence between atoms (a change of length of the total spin vector $\aver{|\vec{S}|}$), from dephasing between the atomic spin vector $\vec{S}$ and the LO. The latter dephasing is dominated by LO phase noise and is the signal we would like to measure (and feed back) when operating the system as a full atomic clock. On the other hand, processes that induce loss of coherence between atoms are, among others, magnetic-field gradients, atom-atom collisions, light scattering by the atoms etc.

The length of the atomic spin vector $\aver{|\vec{S}|}$ can be determined even in the limit where the LO phase noise is much larger than $2\pi$. If we perform many measurements with homogeneously randomly distributed Ramsey phases $\phi_R$ the variance of the variable $S_z{=}\aver{|\vec{S}|}\cos(\phi_R)$ is
$$
\var{S_z}=\frac{\aver{|\vec{S}|}^2}{2\pi} \int_0 ^{2\pi}\cos^2(\phi_R)\mathrm{d}\phi_R = \frac{\aver{|\vec{S}|}^2}{2} 
$$
We can thus infer the mean spin vector length $\aver{|\vec{S}|}$ by measuring the variance of the outcomes of a Ramsey sequence.

\subsection*{Local Oscillator laser}
The local oscillator (LO) laser deployed here is a distributed feedback (DFB) laser diode operating at a wavelength of 1157 nm. 
It uses a fiber-coupled EOM inside the optical feedback path \cite{yamoah2019robust}, allowing to control the laser frequency with high gain and a large bandwidth.
The primary laser is pre-stabilized to a highly-stable, high-finesse optical cavity using
a standard Pound-Drever-Hall frequency stabilization scheme.
The light is used to inject a gain chip to increase the power, and the output of the secondary laser is frequency doubled to reach the optical clock transition wavelength of 578~nm.

\subsection*{LO noise}
The local oscillator noise is characterized by interrogating directly the atoms.
We perform a standard Ramsey sequence every 4~seconds.
We measure the accumulated phase noise variance $(\Delta\phi)^2$ for different Ramsey times $\tau_R$.

We first find that the noise is uncorrelated between different sequences.
This is based on a Ljung-Box test performed after removal of long-time drifts ($>30$~s).
Moreover, we observe that the phase variance scales quadratically with $\tau_R$. For $T_C \gg \tau_R$ we model the atom-LO phase variance as
\begin{equation}
    (\Delta\phi)^2=(\Delta\omega)^2\tau_R^2+\frac{\xi_W^2}{N},
\end{equation}
where $(\Delta\omega)^2$ is the sequence-to-sequence angular frequency variance of the LO, and $\xi_W^2/N$ is the total projection noise induced by $N$ atoms.
With a CSS input, our system has a Wineland parameter $\xi_W^2=1.35$~(see Fig.~\ref{fig:squeezingSurvival} in the main text).
By fitting the data to the model we obtain $\Delta\omega = 2\pi\times(78\pm3)$~Hz.
With the cycle time $T_C=4$~s of our system, we infer a LO short-term fractional stability of $\sigma_{LO} = 3\times10^{-13}~\mathrm{s^{-1/2}}/\sqrt{T}$.

For a Ramsey time $\tau_R=1.16~$ms, a CSS Wineland parameter $\xi_W^2=1.35$, and $N=300$, the projection noise contribution to the fractional stability is $\sigma_{PN}= 3\times10^{-14}~\mathrm{s^{-1/2}}/\sqrt{T}$, and thus negligible compared to LO noise contribution.
Moreover, for a Ramsey time $\tau_R=0.17~\mathrm{\mu s}$ and same atom number, we have $\sigma_{PN}{\approx}\sigma_{LO}$.

We define the LO coherence time $\tau_{LO}$ as the Ramsey time for which the LO noise standard deviation reaches $\Delta\phi(\tau_{LO})=\pi/2$.
This results in $\tau_{LO}\equiv(\pi/2)/\Delta\omega=6~$ms for our current system.

\section*{}

\end{document}